\documentstyle[aps,psfig,epsf]{revtex}

\begin{document}
\title{Huge oxygen isotope effect on local lattice fluctuations\\ 
in La$_{2-x}$Sr$_{x}$CuO$_{4}$ superconductor}

\author{A. Lanzara$^{1}$, N. L. Saini$^{1}$, A. Bianconi$^{1}$,\\
Guo-meng Zhao$^{2}$, K. Conder$^{2}$, 
H. Keller$^{2}$, K. A. M\"uller$^{2}$}
\address{$^{1}$Unita' INFM and Dipartimento di Fisica, Universita' di Roma "La Sapienza'',  P.le Aldo Moro 2, 00185, Roma, Italy \\
$^{2}$Physik-Institut der Universit\"at Z\"urich,
CH-8057 Z\"urich, Switzerland}

\maketitle
\widetext
\vspace{0.5cm}
\begin{abstract}
Recently a growing number of experiments have provided indications of the key role of polarons (composite particles formed by a charge strongly coupled with a local lattice deformation) in doped perovskites, hosting colossal magnetoresistance (CMR) and high T$_{c}$ superconductivity \cite{Goodenough,Muller,Bianconi,Mihailovic,Alexandrov,Xiang,Calvani,Zhao,Chen,Radaelli,Lanzara,Medarde}. While the role of polarons is generally recognized in manganites due to the large amplitude of the local lattice deformation, the scientific debate remains open on the cuprates where the lattice deformation is smaller and the anomalous normal metallic phase becomes complex due to the coexistence of polarons with itinerant carriers. Moreover the segregation of polarons and itinerant charges in different spatial domains forming lattice-charge stripes \cite{Goodenough,Muller,Bianconi} as well as the slow dynamic 1D spin fluctuations \cite{Mook} have been observed. The debate on the driving force for the stripe formation remains object of discussion since it could be purely due to electronic interactions and/or due to strong electron-lattice (polaronic) interactions. In order to explore the important role of the later in stripe charge segregation, we have studied isotope effects on the dynamical lattice fluctuations and polaron ordering temperature.  Here we report a compelling evidence for a huge isotope effect on local lattice fluctuations of La$_{2-x}$Sr$_{x}$CuO$_{4}$ high T$_{c}$ superconductor by x-ray absorption spectroscopy, a fast ($\sim$ 10$^{-15}$ sec) and local probe ($\sim$ 5 $\AA$). Upon replacing $^{16}$O with $^{18}$O, the characteristic temperature T$^{*}$ for polaron ordering in La$_{1.94}$Sr$_{0.06}$CuO$_{4}$ increases from about 110 K to 170 K.
\end{abstract}
\vspace{1.5cm}
Until now most of the proposed theories for high T$_{c}$ superconductivity have considered spin and charge fluctuations in a doped 2D antiferromagnetic plane disregarding lattice instabilities. However, the isotope effect experiments do not support this point of view. In particular, the isotope effects in high-temperature superconductors (HTSC) are very unusual and even larger than that for conventional phonon-mediated superconductors while the HTSC materials are underdoped \cite{Crawford,Franck,Zhao2,Zhao3,Zhao4}. Zhao and co-workers have studied the oxygen isotope effects on both the penetration depth and the carrier concentration in some cuprate superconductors \cite{Zhao2,Zhao3,Zhao4}. These results show that the effective supercarrier mass depends strongly on the oxygen isotope mass in underdoped cuprates, suggesting that polaronic and/or bipolaronic charge carriers exist in HTSC. Although the isotope-effect experiments indicate that lattice excitations are indeed important in the HTSC, some other crucial experiments have to be explained in a consistent way to understand the pairing mechanism. This depends on understanding the complex normal phase, that is a particular phase of condensed matter in which charge segregation occurs in stripes at low temperatures \cite{Goodenough,Muller,Bianconi}. Thus, if lattice excitations are strongly coupled to the charge carriers, the isotope mass should influence the local lattice deformation and the charge segregation temperature T$^{*}$ should depend on the isotope mass.\\
Here we report studies of how the dynamic local lattice distortions and the charge segregation temperature change with the oxygen isotope mass in La$_{1.94}$Sr$_{0.06}$CuO$_{4}$ from x-ray absorption near edge structure (XANES) measurements. We choose this compound where the largest oxygen isotope effects on both superconducting transition temperature and the effective supercarrier mass have been observed \cite{Zhao4}. The XANES spectroscopy probes the statistical distribution of the conformations of the Cu-O cluster via electron scattering by oxygen atoms neighbouring to the Cu within a measuring time of the order of 10$^{-15}$ sec \cite{Bianconi2}. Therefore it is very sensitive to distortions of the local structure of oxygens around the photoabsorbing atom and probes instantaneous lattice conformations. Our results show that the oxygen isotope substitution modifies the local structure of the CuO$_{2}$ plane and shifts the temperature T$^{*}$ for polaron ordering towards a higher temperature by 60 K $i.e.$, 110($\pm$ 10) K in the $^{16}$O sample and 170($\pm$ 10) K in the $^{18}$O sample.
The experiments were performed on the $^{16}$O and $^{18}$O isotope samples at the European Synchrotron Radiation (ESRF), Grenoble. Fig. 1a shows the normalized Cu K-edge x-ray absorption near edge spectra for the $^{16}$O and $^{18}$O samples at 180 K. We denote the well-resolved peak features by A$_{1}$, A$_{2}$, B$_{1}$ and B$_{2}$. These features, which extend from about 8 eV to 40 eV above the threshold, have been  identified as "shape resonances" determined by multiple scattering of the excited photoelectrons scattered with neighbouring atoms, as revealed by XANES calculations for La$_{2}$CuO$_{4}$ \cite{Li}. The resonances B$_{1}$, B$_{2}$ (A$_{1}$, A$_{2}$) are determined by electron scattering with in (out of) plane atoms, therefore the main peak B$_{1}$ probes the multiple scattering from the oxygen atoms in the CuO$_{2}$ square planes while the peaks A$_{1}$, A$_{2}$ are mainly due to the multiple scattering of the ejected photoelectron from the apical oxygen and La/Sr atoms. In the presence of local lattice fluctuations, with a time scale slower than the XANES measuring time, XANES probes the statistical distribution of the Cu sites.
The effect of isotope substitution on the distribution of local Cu site conformations can be directly seen in Fig. 1,  where we report the difference between the spectra of the two isotopes at 200 K. There is a large variation of the XANES spectrum due to isotope substitution, of the order of 2$\%$ of the normalised absorption. This is a direct measure of the effect of isotope mass on the distribution of Cu site structural conformations spanned during the dynamic lattice fluctuations associated with the  polarons. We observe an energy shift of peaks A$_{1}$ (and A$_{2}$), and an intensity decrease of peak B$_{1}$ (and B$_{2}$) in the $^{18}$O spectrum. These changes are clearly related with the effect of changing the oxygen mass on polaronic lattice fluctuations involving the tilting of apical oxygens (out of plane) and rhombic distortion of the Cu-O square planes \cite{Bianconi}.
In order to study the evolution of the polaronic fluctuations as a function of temperature we report the absorption differences (with respect to spectra at 60 K) for the $^{16}$O (Fig.  2a) and $^{18}$O (Fig.  2b) samples. For both samples, the difference spectra show a sudden change at a characteristic temperature T$^{*}$. This characteristic temperature T$^{*}$ is 110($\pm$ 10)K for the $^{16}$O sample, and 170 ($\pm$10)K for the $^{18}$O sample. The oxygen isotope shift on T$^{*}$ is quite large (60 K).\\
We have taken as a conformational parameter the intensity ratio R$=$ $(B_{1}-A_{1})/(B_{1}+A_{1})$ that is plotted in Fig. 3. By decreasing the temperature the ratio R shows an increase at T$^{*}$ where the local lattice fluctuations slow down \cite{Saini} at the temperature for the polaron ordering in lattice-charge stripes \cite{Bianconi}. The temperature dependence of the R (Fig. 3) shows that this crossover temperature T$^{*}$ increases by about 60 K upon replacing $^{16}$O with $^{18}$O. This result indicates that the isotope substitution leads to a large change in the polaronic lattice fluctuations. Incidentally a large isotope effect has been theoretically predicted recently on polaronic lattice modes by Mustre de Leon $et  al$ \cite{Mustre}. 
The T$^{*}$ is associated with the temperature below which the coexisting polaronic and itinerant charge carriers spatially distributed in stripes could be well distinguished within the time scale of the x-ray absorption technique as revealed by atomic pair distribution of CuO$_{2}$ plane \cite{Bianconi,Saini}. When the temperature is lowered, with the fixed time scale of the XANES, measuring change in higher order pair distribution, we observe an anomalous change in the local lattice across the temperature T$^{*}$ due to a sudden change in the statistical distribution of the Cu-site due to a transition in an ordered phase. Obviously, the T$^{*}$ could be different if measured by other techniques depending on the time scale. 
It is worth noting that T$^{*}$ of the virgin samples ($i.e.$, the sample with the sample oxygen isotope) is nearly independent of doping (T$^{*}$ $\sim$110 K for $x$=0.06, T$^{*}$ $\sim$100 K for $x$=0.15, and T$^{*}$ $\sim$150 K for La$_{2}$CuO$_{4.1}$) \cite{Bianconi,Lanzara2}. This indicates that the observed giant oxygen isotope shift of T$^{*}$ is not caused by any possible difference in the carrier concentrations of the two isotope samples. As a matter of fact, oxygen isotope substitution hardly affects the carrier concentrations \cite{Zhao2,Zhao3,Zhao4}.
From Fig. 3, one can also see that below T$^{*}$ the difference of the two isotope samples is much smaller than that above T$^{*}$. This may be explained by the fact that the average local structure deviation in the ordered striped phase due to isotope substitution is less than that in the disordered phase.\\
The large oxygen isotope shift suggests that the stripe formation in high-temperature superconductors is not caused by purely electronic interactions. It appears that the electron-phonon interaction plays an important role in this phenomenon. This is consistent with the large oxygen isotope effects on both T$_{c}$ and the effective supercarrier mass observed in this material \cite{Zhao4}. The large oxygen isotope effect on the supercarrier mass implies that the nature of charge carriers in this material is of polaron and/or bipolaron type. Since the (bi)polaronic carrier for the $^{18}$O sample is heavier than for the $^{16}$O sample \cite{Alexandrov2}, one expects that the average local structure deviation of the $^{18}$O sample should be larger for the given time scale, in agreement with the results shown in Fig. 1 and Fig. 3. The present experiment places important constraints on the microscopic origin of the stripe phase and on the microscopic mechanism of high-temperature superconductivity.

{\bf Methods}

{\bf Sample preparation and properties.} Samples of  La$_{1.94}$Sr$_{0.06}$CuO$_{4}$ were prepared by a conventional solid-state reaction using dried La$_{2}$O$_{3}$ (99.99$\%$), SrCO$_{3}$ (99.999$\%$) and CuO (99.999 $\%$). The powders were mixed, ground thoroughly, pressed into pellets  and fired in air at 1000$^{\circ}$C for $\sim$96 h with three intermediate grindings. The diffusion was carried out for 40 h at 900 $^{\circ}$C and oxygen pressure of about 1 bar. The cooling time to room temperature was $\sim$ 4 h. The oxygen isotope-enrichment was determined from the weight changes of the $^{16}$O and $^{18}$O samples. The $^{18}$O samples had $\sim$ 90$\%$ $^{18}$O and $\sim$10$\%$ $^{16}$O. The $^{16}$O sample has a T$_{c}$ of  about 8 K, and the $^{18}$O has a T$_{c}$ lower by about 1 K than the $^{16}$O sample \cite{Zhao4}.

{\bf X-ray absorption measurements.} The Cu K-edge absorption measurements were performed on powder samples. The absorption spectra were recorded on the beam line BM29 at ESRF, Grenoble. The Cu K$\alpha$ fluorescence yield (FY) off the samples was collected using multi-element Ge element X-ray detector array, covering a large solid angle, to measure the absorption signal. The samples were mounted in a closed cycle He refrigerator and the temperature was monitored with an accuracy of $\pm$ 1 K. To measure the spectra with a very high signal to noise ratio and with a high resolution at the Cu K-edge, we used Si(311) double crystal monochromator and collected several absorption scans at the same temperature. The measurements were repeated on another beam line (BM32) after several months and in spite of different experimental set-ups for the two experimental runs, the results showed a very good reproducibility.

\newpage

{\bf Figure captions:}
Fig. 1: Cu K-edge X-ray absorption near edge structure (XANES) of La$_{1.94}$Sr$_{0.06}$CuO$_{4}$ for the $^{16}$O and $^{18}$O isotope samples (panel a) and their difference (panel b) at 200K. The $^{16}$O$-->$ $^{18}$O isotope effect on the Cu site structure fluctuations induces a decrease of the peak B$_{1}$ with an increase of peak A$_{1}$.

Fig.  2: Temperature dependence of the absorption difference spectra for the two isotopes with respect to the spectra at 60 K for the $^{16}$O (panel a) and $^{18}$O sample (panel b). In both cases the temperature, where the local structure displays a change,  is underlined. 

Fig. 3: Temperature evolution of the XANES intensity ratio R$=$ $(B_{1}-A_{1})/(B_{1}+A_{1})$ where B1(A1) is the intensity of XANES peak B$_{1}$ (A$_{1}$) for the $^{16}$O (upper) and $^{18}$O (lower) substituted samples. There is a sudden change of the intensity ratio at a characteristic temperature T$^{*}$ $\sim$110($\pm$10)K for the $^{16}$O sample, and $\sim$170 ($\pm$10)K for the $^{18}$O sample giving a huge oxygen isotope shift of $\sim$60 K.

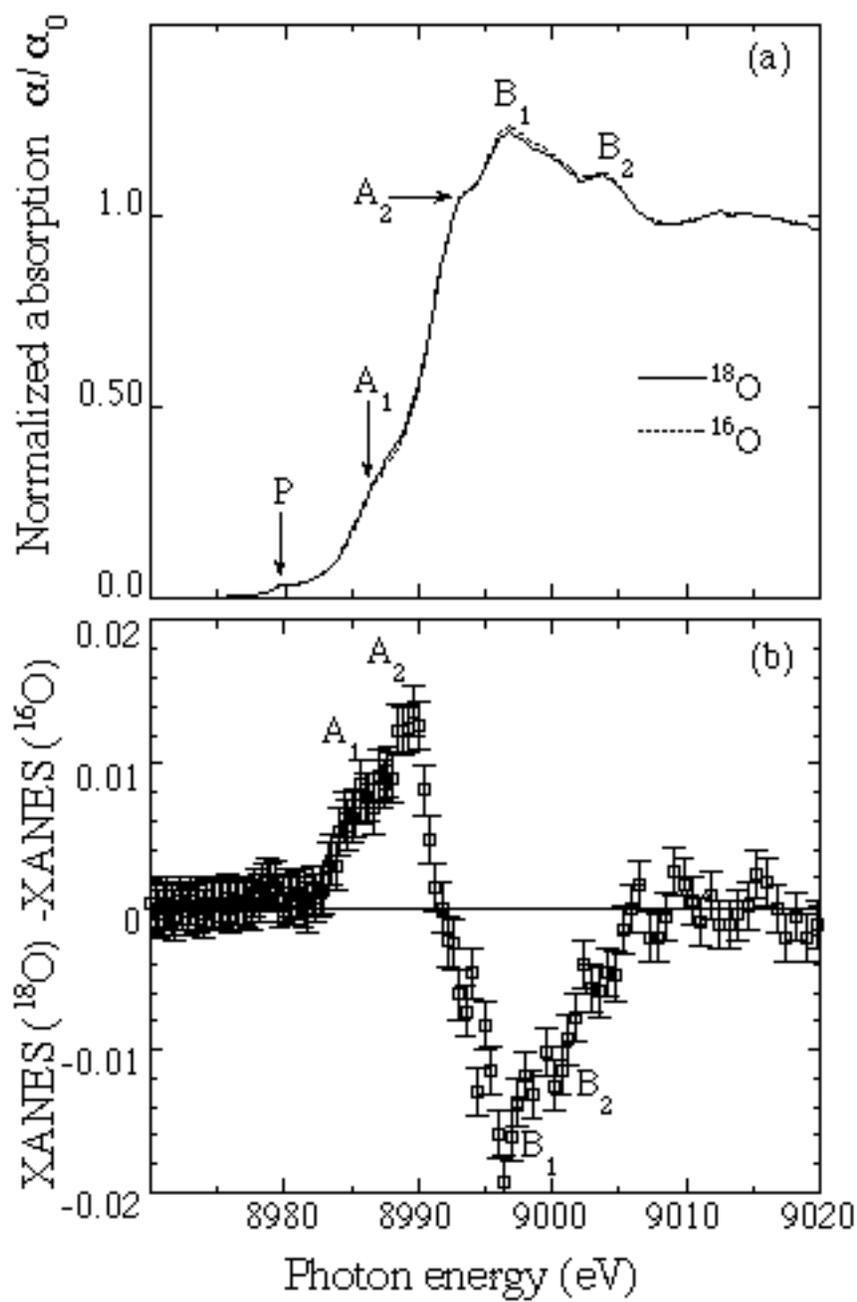
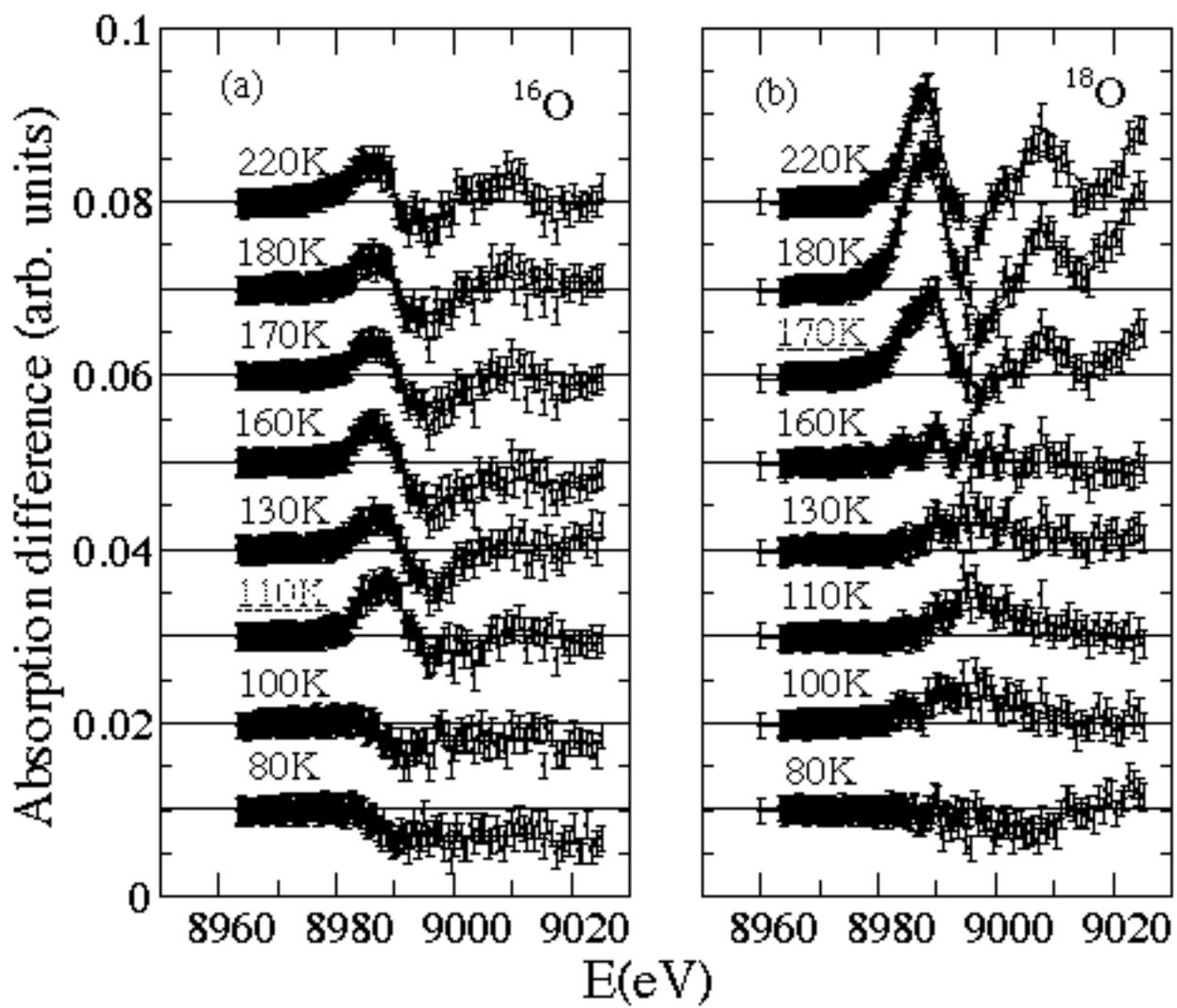
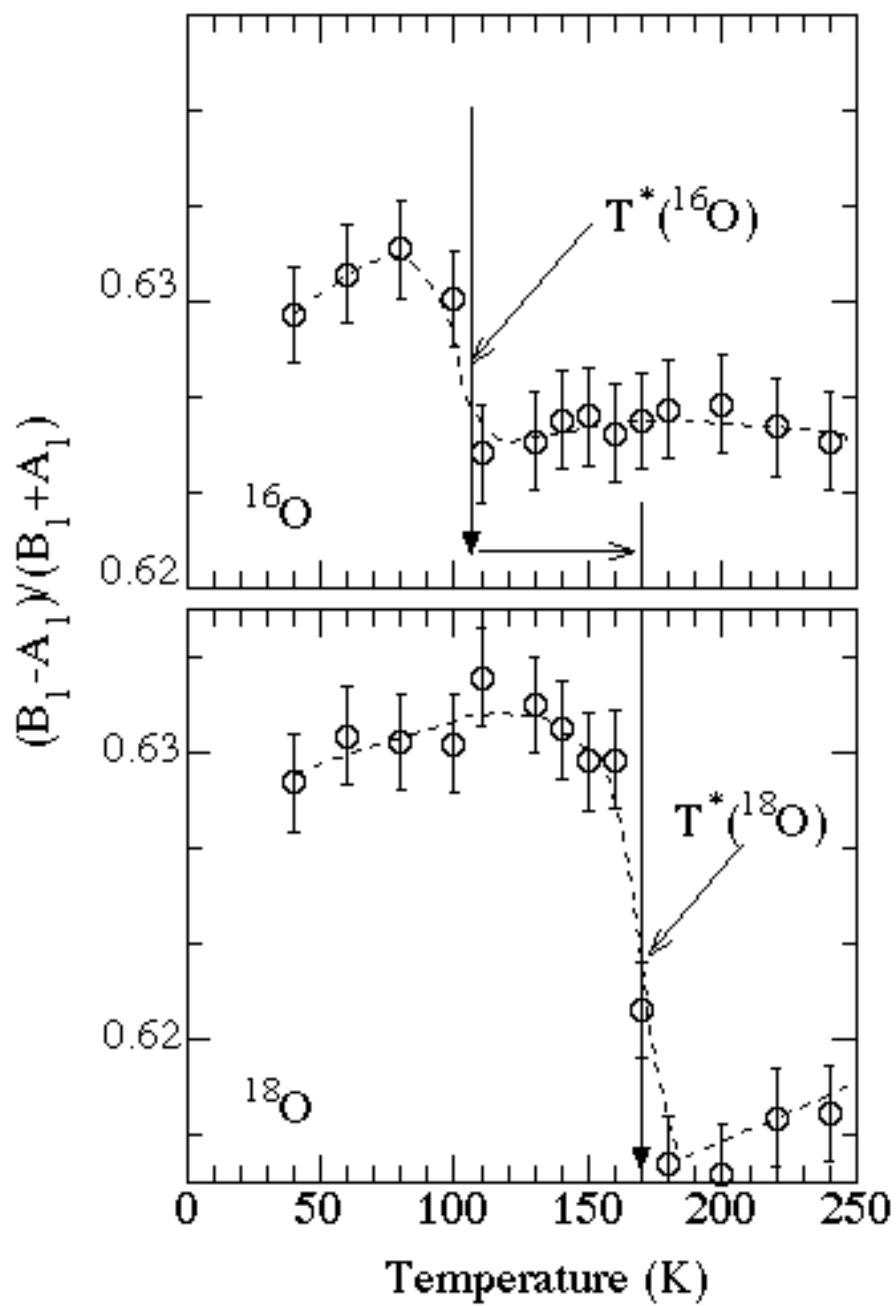
\end{document}